\documentclass[aps,prl,twocolumn,showpacs,preprintnumbers,amsmath,amssymb,groupedaddress,superscriptaddress]{revtex4-1}

\usepackage{graphicx}
\usepackage{dcolumn}
\usepackage{bm}
\usepackage{subfigure}
\usepackage{amsthm}

\begin{document}

\preprint{APS/123-QED}

\title{Frequency adjustment and synchrony in networks of delayed pulse coupled oscillators}

\author{Joel Nishimura}
\affiliation{Division of Mathematical \& Natural Sciences, Arizona State Univeristy, Glendale, Arizona 85306, USA}

\date{\today}

\begin{abstract}
We introduce a system of pulse coupled oscillators that can change both their phases and frequencies; and prove that when there is a separation of time scales between phase and frequency adjustment the system converges to exact synchrony on strongly connected graphs with time delays. The analysis involves decomposing the network into a forest of tree-like structures that capture causality. Furthermore, we provide a lower bound for the size of the basin of attraction with immediate implications for empirical networks and random graph models. These results provide a robust method of sensor net synchronization as well as demonstrate a new avenue of possible pulse coupled oscillator research.
\end{abstract}

\maketitle

Pulse coupled oscillators (PCOs) have proven themselves an incredibly successful model of temporal coordination.  Whether in biological, engineering or physical systems, the mix of discrete and continuous elements in PCO models allow for a detailed study of synchronization in a surprisingly parsimonious and well motivated system \cite{Strogatz}. 

One measure of the success of pulse coupled oscillator synchronization is its adoption for a family of wireless sensor network synchronization protocols \cite{Anna,pagliari2011scalable,wang2012energy}.   However, while traditional PCO models provide an excellent tool to study synchronization in idealized settings or with specified network topologies, its application to wireless sensor networks has revealed that when such idealized PCOs are generalized to more realistic settings, they typically have great difficulty synchronizing.  In particular, traditional PCO models are especially challenged by the combination of complex network topologies and signal delay \cite{KlinglmayrStochastic, Timme, secondPaper, firstpaper}; this has naturally led to a number of design questions relevant to both those interested in superior wireless sensor network synchronization protocols and those interested in the theoretical limits of the PCO framework.  

The design challenge posed by complex network topology and delays has been recently addressed by a variety of specialized PCO models which augment oscillators with: mixtures of inhibition and excitation \cite{KlinglmayrStochastic, Timme, secondPaper, firstpaper}, stochasticity \cite{KlinglmayrStochastic}, single bits of addition memory \cite{KeBo,Hu} or other modifications \cite{Nakada2012}. These recent PCO models represent a surprisingly large break from traditional PCO studies and from dynamical systems more generally---requiring new analytical techniques, new theoretical goals and new considerations for novelty. 

However, while these new models have dealt with very difficult settings, they have been unable to address one of the more interesting traditional oscillator questions: can oscillators with heterogeneous frequencies synchronize?  Of the PCO models able to synchronize on a complex network with delays,
there is at best numerical evidence that they approximate synchrony when oscillator frequencies are heterogeneous.  In these more complicated settings, there is little understanding of how to design PCO systems to handle heterogeneous frequencies---and under reasonable assumptions, exact synchrony is clearly impossible.  Yet, given that frequency alteration is common in Hebbian Learning \cite{righetti2006dynamic} and the recent advances in  continuous oscillators \cite{mallada2011distributed}, a PCO model which allows individual oscillators to adjust their frequencies is not only well motivated, but promising.

As such, the main contribution of this paper is the introduction of a system of phase-frequency pulse coupled oscillators and proof that this system attains exact synchrony even in the presence of time delays, complex connected networks, and oscillator frequency heterogeneity.
This definitively answers the question of whether exact synchrony is possible in such systems.

 In particular, we show that when there is a separation in time scales between phase and frequency changes there is an invariant cascading region of phase-frequency space and a corresponding phase-locked fixed point.  To build the machinery for this result, we analyze a single oscillator subjected to periodic forcing and oscillator pairs. Interestingly, these simplified results are then conglomerated via an analysis of the emergent tree-like dependencies in the system, yielding the main convergence result for connected networks. Subsequently, at this fast-time fixed point, the slow-time frequency responses drive the system towards exact synchrony. Next, we investigate the basin of attraction for this oscillator system, yielding a lower bound on the probability of convergence based on an analysis of the networks' degree sequence.  Finally, to bolster these analytic results, we provide numerical simulation demonstrating the robustness of the system.

Similar to \cite{ secondPaper, firstpaper, guardiola2000synchronization,diaz1998mechanisms}, consider a PCO model on an undirected graph $G  = \{V,E\}$, where each oscillator $i\in V$ has phase $\phi_i(t) \in [0,1]$ and speed $\omega_i \in [1,2)$ such that $\frac{d \phi_i}{dt} = \omega_i$. When $i$ reaches its terminal phase, $\phi_i(t) = 1$, it emits a signal and its phase is reset to $0$. The signal from $i$ takes time $\tau<\frac{1}{4}$ to $i$'s neighbors in $G$: $N(i)$. Let $t^n_i$ denote the time of the $n$'th firing of oscillator $i$.  We limit the rate that oscillators can respond to incoming signals by introducing a `quiescent' period, where after an oscillator $j$ processes an incoming signal, it then ignores future signals for the next $q>2\tau$ time.  Otherwise, when a signal from $i$ arrives at non-quiescent oscillator $j$, $j$ adjusts both its phase and frequency according to its phase resetting curve, $f$ and frequency response curve $g$.  Namely:  $\phi_j(t_i^n+\tau) \gets f( \phi_j(t_i^n+\tau) )$ and $\omega_j(t_i^n+\tau) \gets \omega_j(t_i^n+\tau) [1+\epsilon g(\phi_j(t_i^n+\tau))]$ for small $\epsilon>0$.  

We consider phase resetting curves and frequency response curves of the form: 
\begin{displaymath}
   f(\phi) = \left\{
\hspace{-1mm}
     \begin{array}{cl}
     (1-\alpha) \phi  & : \phi <B \\
     1 & : \phi\ge B
     \end{array}
\right.
\Bigg| \quad
   g(\phi) = \left\{
\hspace{-1mm}
     \begin{array}{cl}
	0    & : \phi \in [0,\tau) \\
     <0  & : \phi \in [\tau,B) \\
     \ge 0  & : \phi\ge B
     \end{array}
\right.
\end{displaymath}
with parameters $\frac{1}{2} < \alpha < 1$ and  $B\le \frac{1}{2} -2\tau$.

It is worth highlighting that this model expands on models used in \cite{firstpaper,secondPaper,guardiola2000synchronization,diaz1998mechanisms} in two important ways.  The first way this model differs is that it allows oscillators to adjust their frequency via a frequency response curve---this allows oscillators to overcome heterogeneity in oscillator frequency.  The second modification is the introduction of the quiescent period.  The quiescent period operates analogously to, but is quite different than, the well studied refractory period \cite{KlinglmayrStochastic,Konishi}. Conceptually, the quiescent period corresponds to a situation where oscillator receptors are overloaded by processing an incoming signal, such that they can only process the first of a series of closely timed signals. 

We now consider the slow time subsystem, where phases change but not frequencies.  First, consider an oscillator $i$ that receives a time-delayed periodic forcing at times $t^n_{\hat{\omega}}$ with frequency $\hat{\omega}$,  ($ \frac{\omega_i}{B} \ge \hat{\omega} \ge \frac {\omega_i}{1 +2 \alpha \tau \omega_i}$). If $\hat{\omega}>\omega_i$ and $i$ were to phase lock to the forcing, then one would expect $i$ to be regularly excited by the faster periodic forcing---in fact this  happens.  Specifically, at the phase-locked fixed point, $i$ would fire each time the periodic signal arrives, at times $t_i^{n+1} = t^n_{\hat{\omega}} +\tau$. 

The more interesting case is when $\hat{\omega}<\omega_i$.  In this case, $i$ must be inhibited if it is to phase-lock with the periodic signal.  When $i$ is inhibited,  $\phi_i(t_{\hat{\omega}}^n+\tau) \gets  (1-\alpha) \phi_i(t_{\hat{\omega}}^n+\tau)$, and this leads it's next firing  to come before the periodic forcing, at time: $t_i^{n+1} = \frac{1}{\omega_i} + \alpha(\tau+t_{\hat{\omega}}^n- t_i^n)$.  A slightly more detailed analysis of these two cases concludes that $t_i^{n} - t_{\hat{\omega}}^n $ converges to $\min \{ \tau, \tau + \frac{\hat{\omega} - \omega_i}{\alpha \hat{\omega} \omega} \}$.

Extending the above analysis gives that if an oscillator $i$ processes exactly one signal in $[t_i^{n},t_i^{n+1})$ then either: 
\begin{equation}
t_i^{n+1} = \frac{1}{\omega_i} + \alpha(\tau-\Delta t_i^n)
\label{eqnTiming}
\end{equation}
where $\Delta t_i^n = t_i^n -\min_{j \in N(i)} t^n_j$, or $i$ gets excited to firing by some oscillator  $k$, leading to $t_i^{n+1}  = t_k^{n+1} +\tau$.

Now consider a pair of oscillators $(i,j)$ with $\Delta t^n = t_i^n - t_j^n  \le \tau$. Manipulation of Eq. (\ref{eqnTiming}) gives that the evolution of $\Delta t ^{n+1} = \min\{ \tau, \Delta t (1- 2\alpha) +\frac{ \omega_j - \omega_i}{ \omega_i \omega_j}\}$, provided the frequencies are not too dissimilar, $\omega_i \le \omega_j \le \frac{\omega_i}{ B}$.  Thus, a pair of oscillators evolves to fixed point, $\Delta t =\min\{\tau,  \frac{ \omega_j - \omega_i}{2\alpha \omega_i \omega_j} \}$. Eq. (\ref{eqnTiming}) also gives that the average period across the pair $(i,j)$ is $\min \{ \frac{1}{2}(\frac{1}{\omega_i}+\frac{1}{\omega_j}) + \alpha \tau , \frac{1}{\omega_j} + 2\alpha \tau \}$.

\begin{figure}
\centerline{
\subfigure {
\includegraphics[width=.29\textwidth]{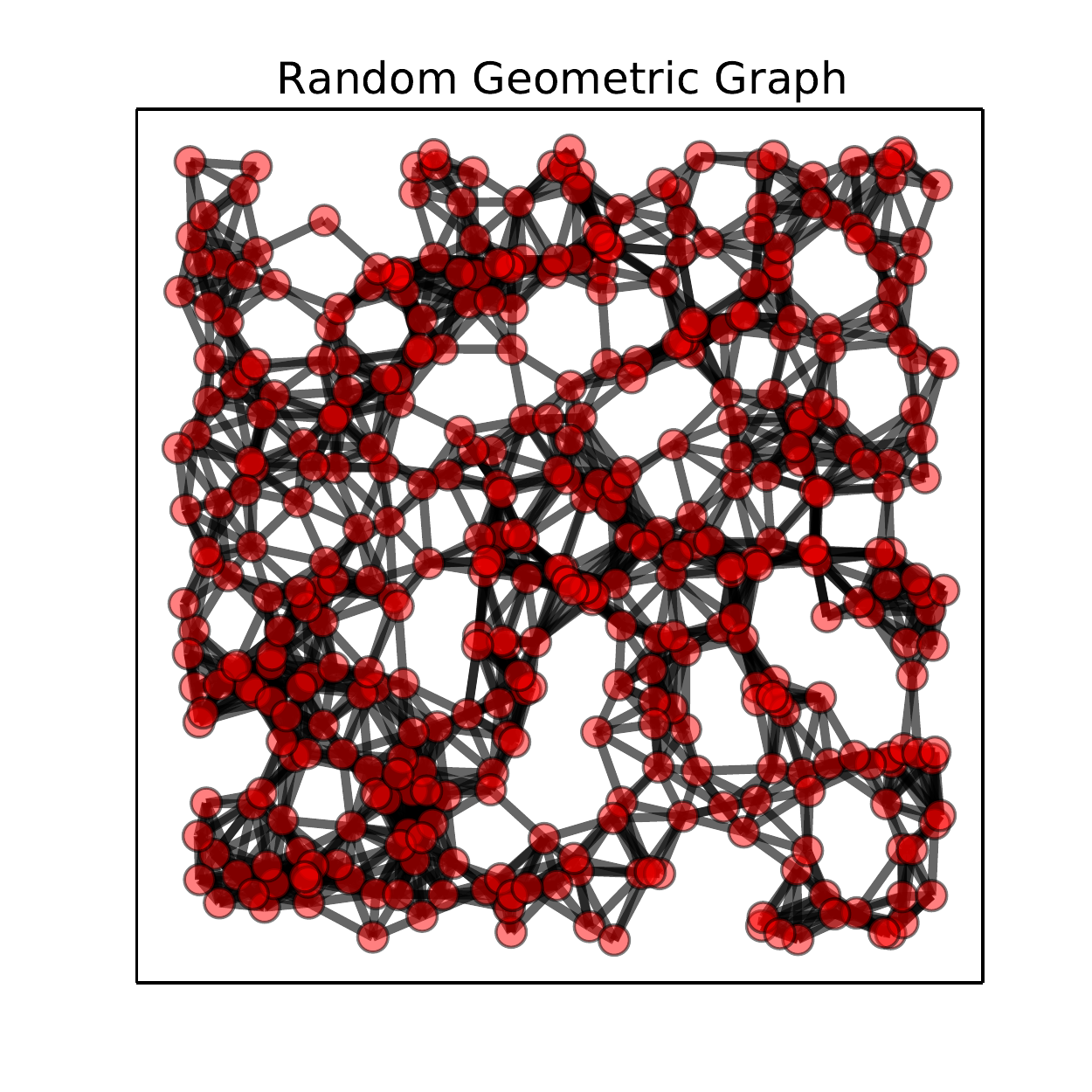} 
} \hspace{-.35 in}
\subfigure {
\includegraphics[width=.29\textwidth]{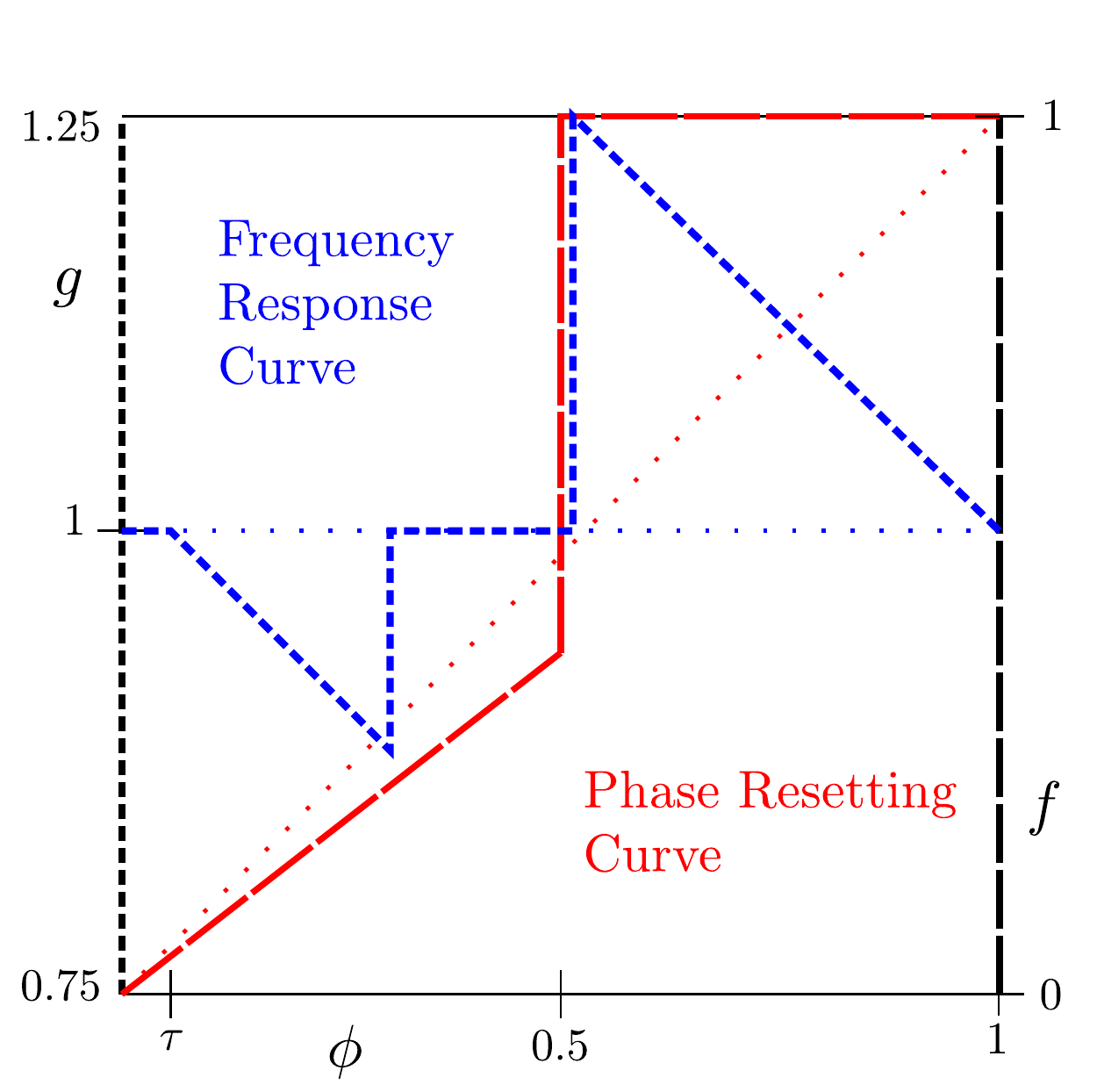} 
}}
\vspace{-3 mm}
\centerline{
\subfigure {
 \includegraphics[width=.29\textwidth]{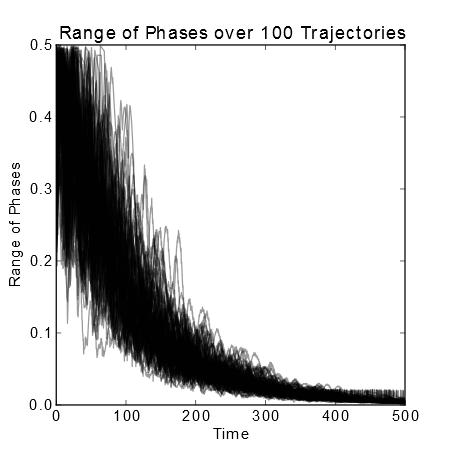}
} \hspace{-.35 in}
\subfigure {
\includegraphics[width=.29\textwidth]{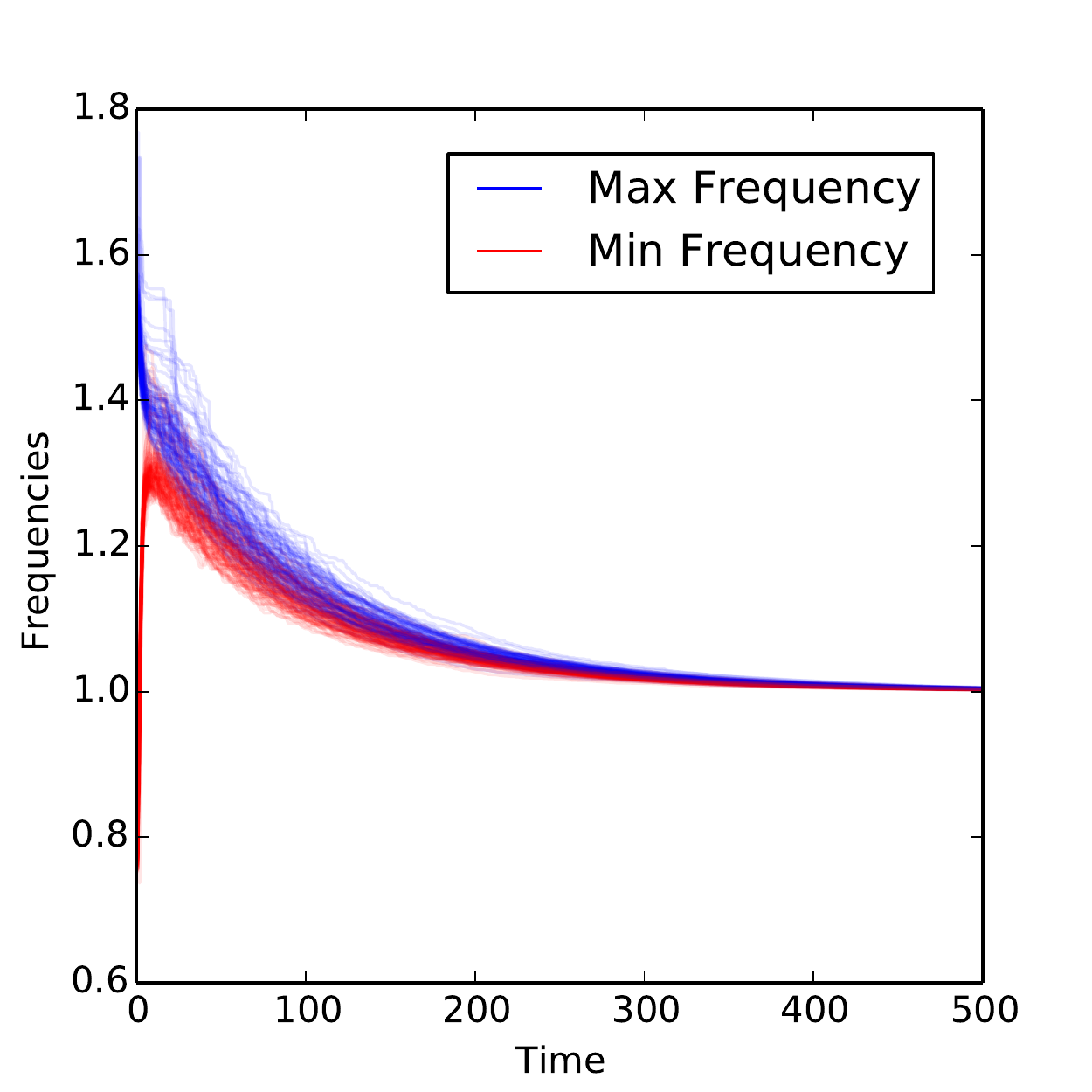}
}}
\vspace{-3 mm}
\centerline{
\subfigure {
 \includegraphics[width=.29\textwidth]{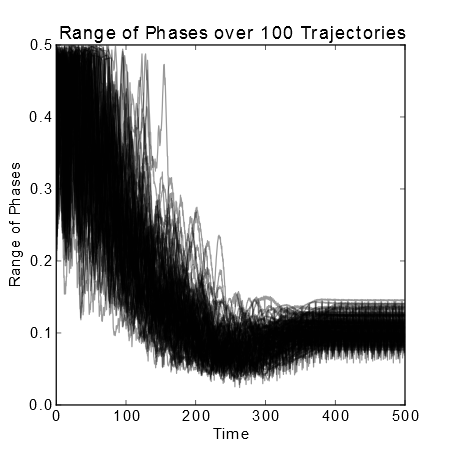}
} \hspace{-.35 in}
\subfigure {
\includegraphics[width=.29\textwidth]{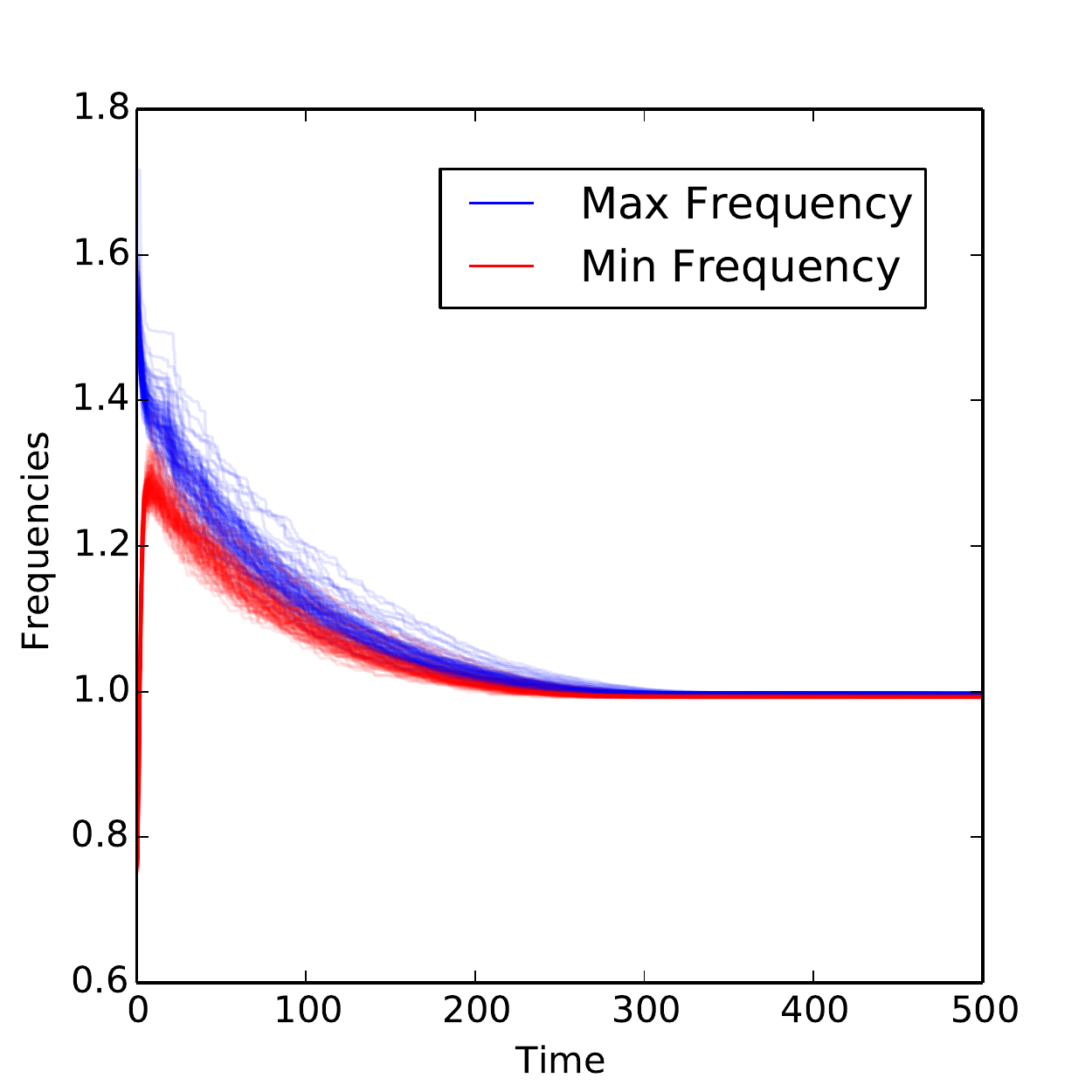}
}}
\vspace{-5 mm}
\caption{(color online) Numerical simulations of $100$ oscillators on a random geometric graph (top left) with given phase resetting curve and frequency response curve (top right).  (middle) Both frequencies and phases converge to synchrony even when the frequency response is not small. (bottom) The method remains robust to heterogeneous delays, where frequencies converge, but phases do not. }
\label{numericalConv}
\end{figure}

As we will show, the simplified behavior due to a pair of oscillators and constant forcing can be aggregated to describe the behavior of the overall fast system with static, but heterogeneous frequencies. 

First, we show that the extremely excitatory tail of the phase resetting curve creates an invariant region of phase space.  Let the finite cascading region  be the portion of phase space where the n'th time each oscillator fires it does so within $\tau$ time of its neighbors, i.e. $|t_i^n - t_j^n|\le \tau$ for all $(i,j) \in E$.  In most cases, this finite cascading region is invariant for a simple reason: when an oscillator's signal arrives at it's neighbor $\tau$ time later, it either then forces that oscillator to fire, or that oscillator must have already fired.  In either case $|t_i^{n+1} - t_j^{n+1}|\le \tau$. 

Inside of the finite cascading region, notice that while oscillators fire at times $t_i^n$, they must receive a signal by $t_i^n+\tau$, and cannot receive another signal until $t_i^{n+1}\ge t_i^n + \frac{1}{2}-\tau$.  This splits the times between those when oscillators are interacting with signals, and when the oscillators are simply integrating their speed.  While the excitatory tail is vital to ensure that oscillators stay in the finite cascading region, it is not particularly illuminating for the remaining properties.  In many of the following arguments, the case where one oscillator is forced to fire by another oscillator is a sticky special case---though many such cases can be dealt with by simply analyzing a similar system, where the frequency of the oscillator excited to firing is increased so that it fires precisely when the incoming signal would have forced it to fire anyways.  

Another important feature of the finite cascading region is that all oscillators share the same asymptotic frequency, and the effect of the heterogeneous frequencies is to determine the differences between oscillator phases. Ideally, it would be the case that the system dealt with these heterogeneous phases in a roughly assortative way, where faster oscillators fire before slower oscillators. Interestingly, this is not the case. 

Instead, we now show that inside this cascading regime the behavior of pairs of oscillators can be well understood.  First, denote the average time a pair of oscillators fires during the n'th round as:
\begin{equation}
v^n_{i,j} = \frac{1}{2}(t_i^n + t_j^n) \label{vDef}
\end{equation}
for $(i,j) \in E$.  Furthermore, for $(i,j)\in E$ define the innate pair period as $p_{i,j} = \min \{ \frac{1}{2} ( \frac{1}{\omega_i} + \frac{1}{\omega_j} )+\alpha\tau, \frac{1}{\omega_j} + 2\alpha \tau \}$ (notice: this would be the average period of $(i,j)$ if they were isolated).  We will show that the order in which oscillators fire is roughly assortative, not on the values of $\omega_i$, but on the values $p_{i,j}$, where pairs with shorter innate $p_{i,j}$ tend to fire first. We build the intuition for this assortativity by investigating causality inside a single round of firing. 

\begin{figure}
\includegraphics{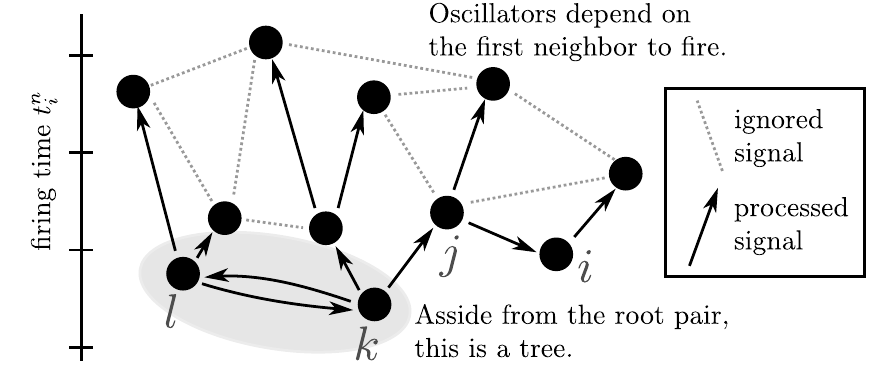}
\vspace{-8mm}
\caption{Since each oscillator processes only a single signal, a round of firing induces a tree-like set of dependencies.  Node labels are consistent with $l \to^n k \to^n j \to^n i$, and $k \to^n l$.}
\label{rootPairTree}
\vspace{-4mm}
\end{figure}

Denote $i$ processing a signal by $j$ in the $n$th round of firings as $j \to^n i$. Notice, $j \to^n i$ is equivalent to the statement 
that $j \in \mathrm{argmin}_{k \in N(i)} t_k^n $ or that $\Delta t_i ^n = t_i^n - t_j ^n$. The quiescent period gives that in a round of firing each oscillator responds to exactly one incoming signal, and thus each round of firings takes the full network of oscillators and creates a directed tree-like structure described by which oscillator depends on which other oscillator.  In this structure, loops are disallowed with one notable exception: the root of each tree must still process exactly one signal from one of its neighbors. Thus, these are trees with the exception that, as seen in Fig. \ref{rootPairTree}, the root is effectively a pair of nodes. Denote the vertex set of the $n$th firing tree with root pair $r$ as $T_r^n$.

We can now begin assembling the pieces we need to show that the system is assortative on $p_{i,j}$.  First, consider oscillators $i$, $j$, $k$ and $l$ (possibly $l=j$), such that $l \to^n k \to^n j \to^n i$ (see Fig. \ref{rootPairTree} for an example of such a chain of oscillators).  Manipulating Eqs. (\ref{eqnTiming}) and (\ref{vDef}) yields that:
$$v^{n+1}_{i,j} = v^n_{i,j}  + p_{i,j}  - \frac{\alpha}{2}( \Delta t_i + \Delta t_j).$$
Since, $k \to^n j \to^n i$  then $\Delta t_i + \Delta t_j  = t_i^n - t_k^n$.  Thus,
$ v^{n+1}_{i,j} = v^n_{i,j}  + p_{i,j}  - \frac{\alpha}{2}( t_i^n - t_k^n)$ and similarly,  $ v^{n+1}_{j,k} = v^n_{j,k}  + p_{j,k}  -  \frac{\alpha}{2}( t_j^n - t_l^n)$. 
Since $k \to^n j \to^n i$ and $l\to^n k \to^n j$, then $t_i^n - t_k^n \ge 0$ and $t_j^n - t_l^n \ge 0$ respectively, yielding:
\begin{equation}
v^{n+1}_{i,j} - v^{n+1}_{j,k}  \ge p_{i,j} - p_{j,k}. 
\label{pairInequality}
\end{equation}
This implies that if $p_{i,j} \ge p_{j,k}$, then $v^{n+1}_{i,j} \ge v^{n+1}_{j,k}$. Thus, if the system is currently assortative on pair periods, it will stay that way.  Moreover, we will show that eventually the first oscillator pair to fire is also one of the fastest pairs.

Applying Eq. (\ref{pairInequality})
repeatedly gives that for any pair $(i,j)$ and root pair $r$,  $v^{n+1}_{i,j} - v^{n+1}_r  \ge p_{i,j} - p_r$.  Consequently, since the root pair evolves independent of the rest of the network with period $p_r$ then, $v^{n+1}_{i,j}   \ge p_{i,j} + v^{n+1}_r- p_r = p_{i,j} + v^{n}_r $.  Combining this over all trees $T_r$ gives that  
$v^{n+1}_{i,j} - v_-^n  \ge p_{i,j} $
where,  $v_-^n = \min_{p,q \in E} v_{p,q}^n$.  Thus, the only possible way that the asymptotic period of the system can equal the minimum pair period, $p_-$, is if there is some pair $(i,j)$, $p_{i,j} = p_-$ and $v^n_{i,j} = v^n_-$. Since the fastest pair $(i,j)$ has the effective period  $v^{n+1}_{i,j}-v^{n}_{i,j} \le p_{i,j}$ then the asymptotic period must come to be $p_{i,j}$, giving that eventually the first nodes to fire are some fastest pair.  Moreover, once this fastest pair becomes a root, no oscillator can come to fire before it.

Assortativity on pair periods allows an understanding of the phase locking fixed point inside the finite cascading region.  First, consider a system with only a single fastest pair, $(i,j)$, at a time when this pair is the first to fire in a round.  Since the signals that $i$ and $j$ process are from each other, the behavior of these oscillators is isolated from the rest of the network, and thus the phase difference evolves to stable fixed point $\min\{\tau,  \frac{ \omega_j - \omega_i}{2\alpha \omega_i \omega_j} \}$. As this lead pair converges to its stable fixed point, it provides a periodic forcing at rate $\hat{\omega} = p_{i,j}$ to its neighbors, whereupon such a neighbor $k$ evolves to a phase difference of  $\min \{ \tau, \tau + \frac{\hat{\omega} - \omega_k}{\alpha \hat{\omega} \omega} \}$ and retransmits the same periodic forcing to its own neighbors.  Thus, the system converges to fixed phase differences which can be computed iteratively out from the fastest pair.  

When there are multiple pairs that all have the same fastest pair period, there are multiple different possible stable equilibrium.  Namely, in these equilibria the graph is divided into several firing trees, each rooted by pairs with the fastest 
 the fastest pairs can each be the root of their own firing trees, with phase differences inside each tree given by the above argument.  

With a clear understanding of this phase locking behavior, we now consider the slow frequency adjustments at that fixed point.  First, let $\omega_+ = \max_{j\in R} \omega_j$, where $R$ is the set of nodes in the root pairs.  Next, consider any root pair of oscillators $(i,j)$, with $\omega_j \ge \omega_i$.  Since the phase difference between $i$ and $j$ is determined by the phase locking difference, it can be shown that $j$ receives its signal from $i$ when:  $\phi_j (t_i^n + \tau) = \omega_j \tau + \frac{\omega_j-\omega_i}{ \alpha\omega_i }$.  When $\omega_i>1$ this implies that $j$ receives a signal when $\phi_j>\tau$, which according to the frequency response curve implies that $\omega_j$ decreases.  Thus, at the phase-locked fixed point the slow frequency response forces the faster oscillator in any root pair to slow down.  

Furthermore, as $\epsilon \to 0$ the remaining oscillators may speed up or slow down, but will not speed down below $1$ or speed up past $\omega_+$.  For example, a forced oscillator $k$ receives it's signal when $\phi_k (\hat{t}^n+\tau) = \frac{\omega_k-\hat{\omega}}{\alpha\hat{\omega}} $.  This is only greater than $\tau$, and thus sped down, when  $\omega_k > \hat{\omega}(1+\alpha \tau) >1$ (provided the root pair has speeds $1$ or greater).  Similarly, for $k$ to be sped up it must be that $k$ is excited to firing, but that requires that $\omega_k < \hat{\omega}$ which implies that $\omega_k <\omega_+$.  Similar analysis can show that the slower oscillator of a root pair $(i,j)$ receives its signal from $j$ when $\phi_i (t_j^n + \tau) = \omega_i \tau - \frac{\omega_j-\omega_i}{ \alpha\omega_i }$, and thus, also stays in the frequency range of $[1,\omega_+]$.

Since all frequencies in the system stay inside $[1,\omega_+]$, and $\omega_+$ is constantly and continuously decreasing then all frequencies must converge to $1$.  The corresponding fixed phase difference at $\omega_i = 1$ for all $i$ is $\phi_i = \phi_j$ for $i$ and $j$.  Thus, as $\epsilon \to 0$, any initial conditions inside the finite cascading region converge to exact synchrony.  

This raises the natural question: what is the probability random initial conditions put the system in the finite cascading region? Utilizing techniques from \cite{secondPaper} yields a lower bound on the probability that the system enters the finite cascading region, and therein synchronizes.  Since the phase resetting curve has a strongly excitatory tail, if every oscillator receives a signal in $t\in (0,t']$, for small $t'$ then every oscillator will either be excited to firing (and thus have small phase) or be inhibited towards $0$ and thus have a small phase.  Once oscillators can be known to all have somewhat small phases, it can be shown that the next batch of firing guarantees that all oscillators are in the finite cascading region.   As a precursor to calculating the probability of this sequence of events, we first determine the critical time $t'$. Namely, if $t'<\frac{\alpha B}{\omega_+}$ then $\phi_i(t' +\tau)<B$ for all $i$.   Similarly, if $t' < \frac{1}{\omega_+}-(\frac{1}{\omega_+}+\frac{1}{\omega_-})B+\frac{\alpha B}{\omega_+}-\tau$ then when the signals from the next batch of firings arrive all the oscillators will be in the excitatory region or have just fired.  The probability that every oscillator fires or receives a signal before time $t' = \min \{\frac{\alpha B}{\omega_+} , \frac{1}{\omega_+}-(\frac{1}{\omega_+}+\frac{1}{\omega_-})B+\frac{\alpha B}{\omega_+}-\tau   \} $ can be calculated using the union bound.  In particular, if the probability that $P(\phi_i(0)\in [ 0, 1-t'+\tau) )= q_i$ then the probability network $G$ with degree sequence $d_i$ converges to the finite cascading region and thus to synchrony is 
\begin{equation}
P_{sync}(G)  \ge 1 - \Sigma_i^n q_i^{d_i+1}.
\label{probSync}
\end{equation}
 Such bounds are known to produce surprisingly strong results \cite{secondPaper}.  For example, such a bound gives a lower bound for a phase transition (from no synchrony to synchrony) on an  Erd\H{o}s-R\'{e}nyi  random graph at only a constant multiple of the critical parameter for percolation.  Similar bounds exists for random geometric graphs and in fact any random graph model which produces predictable degree distributions.  

While Eq. (\ref{probSync}) is extremely effective in bounding the convergence probabilities of large random graphs, Fig. \ref{numericalConv} displays numerical results, showing that the system converges to exact synchrony reliably for intermediate sized systems, and that the system is robust to the inclusion of heterogeneous delays as well as fast frequency response.  

Thus, we have shown how the inclusion of a quiescent period and a frequency response curve in a pulse coupled oscillator system can extend exact synchronization to include delays, complex networks and heterogeneous frequencies.  The analysis builds upon a separation of time scales and an understanding of a pair of oscillators to describe an arbitrary connected undirected network.  This result was then shown to give a bound on the probability of synchronization in large random networks, and the robustness of these results was supported by numerical simulation.  

This research has been supported in part by the NSF under grant CDI-0835706 and the NIH under grant K25 NS-703689-01.

\bibliography{OscReferences5}

\end{document}